\documentclass[aps,prev,twocolumn,preprintnumbers,floatfix,nofootinbib]{revtex4-1}
\usepackage{graphicx}
\usepackage{epstopdf}
\usepackage{mathrsfs}
\usepackage{amssymb}

\usepackage{verbatim}
\usepackage{color}
\usepackage{multirow}

\usepackage{amsmath}
\usepackage{epsfig}
\usepackage{ulem}
\usepackage{graphics}
\usepackage{indentfirst}
\usepackage{slashed}

\newcounter{RomanNumber}
\newcommand{\MyRoman}[1]{\setcounter{RomanNumber}{#1}\Roman{RomanNumber}}
\def\stacksymbols #1#2#3#4{\def\theguybelow{#2}
    \def\vp{\lower#3pt}
    \def\sp{\baselineskip0pt\lineskip#4pt}
    \mathrel{\mathpalette\intermediary#1}}

\def\intermediary#1#2{\vp\vbox{\sp
     \everycr={}\tabskip0pt
     \halign{$\mathsurround0pt#1\hfil##\hfil$\crcr#2\crcr
              \theguybelow\crcr}}}

 \newcommand{\bea}{\begin{eqnarray}}
\newcommand{\eea}{\end{eqnarray}}

\newcommand\nn{\nonumber}

\def\be{\begin{equation}}
\def\ee{\end{equation}}
\def\ba{\begin{eqnarray}}
\def\ea{\end{eqnarray}}

\def\<{\langle}
\def\>{\rangle}

\def\nn{\nonumber}

\def\cO {{\cal O}}

\newcommand{\cL}{\mathcal{L}}
\newcommand{\nc}{\newcommand}
\nc{\LL}{L}
\nc{\vv}{\tilde{v}}
\nc{\ccdot}{\!\cdot\!}
\nc{\gsm}{G_{SM}}
\nc{\vfive}{\mathbf{5}\oplus\mathbf{\overline{5}}}
\nc{\vten}{\mathbf{10}\oplus\mathbf{\overline{10}}}
\nc{\zhol}{Z^{\rm hol}}
\nc{\xfb}{\,{\rm fb}}

\begin{document}

%\wideabs{
%\begin{flushright}
%
%
%\end{flushright}

\preprint{MI-TH-1759}

\vspace*{1mm}

\title{3D CFT Archipelago from Single Correlator Bootstrap}

\author{Zhijin Li$^{a}$}
\email{lizhijin@physics.tamu.edu}
\author{Ning Su$^{b}$}
\email{suning1985@gmail.com}

\vspace{0.1cm}
\affiliation{
${}^a$
George P. and Cynthia W. Mitchell Institute for
Fundamental Physics and Astronomy,
Texas A\&M University, College Station, TX 77843, USA}

 \affiliation{
${}^b$
CAS Key Laboratory of Theoretical Physics, Institute of Theoretical Physics, Chinese
Academy of Sciences, Beijing 100190, China}

\begin{abstract}
 We show that the scaling dimensions of lowest operators in conformal field theories (CFTs) can be isolated in small and closed regions from single correlator bootstrap.  We find the conserved currents play crucial roles in bootstrapping the crossing equation. By imposing a mild gap between the scaling dimensions of the conserved current and its next operator, the scaling dimensions of lowest operators are forced to lie in small isolated regions, i.e., these CFTs can be almost fixed by few lowest operators in certain channels. For CFTs with extended supersymmetry, the single correlator crossing equation involves several conserved or shorted operators and by imposing gaps in these sectors it is possible to isolate different CFTs. Specifically, we bootstrap the isolated regions corresponding to the 3D Ising model, $O(N)$ vector model, $N=1,2$ supersymmetric Ising models  by introducing mild gaps in certain sectors with conserved or shorted operators.
\end{abstract}

\maketitle

%% The arXiv's use of hypertex conflicts with revtex4's use of
%% \tableofcontents in single column format. To avoid this problem,
%% Include a file OOREADME.xxx with the word nohypertex in it when
%% you submit to the arXiv.
%\tableofcontents

\setcounter{equation}{0}

%%%%%%%%%

%%%%%%%%%%%%%%%%%%%%%%%%%%%%%%%%%%%%%%%%%%%%%%%%%%%%%%%%%%%%%%%%%%%%%%

\section{Introduction}
The conformal bootstrap \cite{Polyakov:1974gs, Ferrara:1973yt} which aims to solve the conformal field theories (CFTs) using general consistency conditions, has been revived to study CFTs in higher dimensions ($D>2$) since the seminal work \cite{Rattazzi:2008pe}.
The consistency conditions employed in conformal bootstrap are unitarity and the crossing symmetry of four point correlators. Rigorous bounds on CFT data, including the operator scaling dimensions and operator product expansion (OPE) coefficients,
can be obtained by bootstrapping a single correlation function. The most striking results are obtained from mixed correlators in Ising model and $O(N)$ vector models in $D=3,5$   \cite{Kos:2014bka, Kos:2015mba},  in which the scaling dimensions of lowest scalars can be isolated into small islands. The isolated regions shrink notably with higher order of derivatives and they may converge to unique solutions of specific CFTs.

The key to obtain isolated regions from conformal bootstrap is to find the dynamical constraints besides the general consistency conditions so that one can carve out the targeted CFT. By bootstrapping single correlator, one normally get general bounds on CFT data instead of closed regions, and the conventional wisdom is that the mixed correlators are needed on this purpose, from which we can get access to more channels and impose stronger constraints on the spectra. For the $3D$ Ising model  ($O(N)$ vector models), it shows in \cite{Kos:2014bka, Kos:2015mba} that the isolated regions can be generated by requiring only one relevant $Z_2$ odd scalar and $Z_2$ even scalar (only one relevant scalar in both $O(N)$ vector representations and $O(N)$ singlets).
Nevertheless, for general CFTs, the constraints for isolated regions are usually not so straightforward. In $5D$, the interacting fixed point with $O(N)$ symmetry locates below the free theory and  it needs to introduce more delicate constraints to obtain isolated regions \cite{Li:2016wdp}.
For the supersymmetric CFTs (SCFTs), like 3D supersymmetric Ising model, it gets even more difficult to obtain isolated regions since the fermionic operators play important roles and one may need to bootstrap the mixed correlators with fermions and bosons to fix these SCFTs uniquely.
There is another barrier to apply the mixed correlator bootstrap on many interesting SCFTs: for SCFTs with extended supersymmetry, the mixed correlators contain intricate structures and it is difficult to calculate the superconformal block functions of mixed correlators. Some famous examples are the $3D$ $N=6, 8$ SCFTs and $4D$ $N=2, 4$ SCFTs, which admit fascinating analytical properties and play important roles in AdS/CFT. It would be remarkable if the spectra of these SCFTs can be uniquely determined from conformal bootstrap. However, for these SCFTs, the superconformal partial wave expansions of mixed correlators are generically unknown.

In this work, we study the constraints to obtain isolated sets of solutions on CFT data from conformal bootstrap. Our results suggest that the sufficient conditions are actually concealed in the single correlator.
This conclusion is different from our previous experiences in numerical conformal bootstrap, however, it agrees with recent results from the analytical bootstrap \cite{Gopakumar:2016wkt,Gopakumar:2016cpb,Alday:2016njk,Alday:2016jfr}, in which the perturbative expansions of the CFT data corresponding to the Wilson-Fisher (WF) fixed point can be obtained from the consistency conditions of single correlator. The authors imposed certain constraints implicitly to isolate the WF fixed point from other CFTs. The constraints are hidden in their constructions of higher dimensional operators like $\cO_{2n,\ell}$ based on the fundamental operator $\phi$:
\be
\cO_{2n,\ell}\sim \phi\,\partial^{2n}(\partial_\mu)^\ell\phi.
\ee
The numerical conformal bootstrap is not sensitive to the specific constructions of the higher dimensional operators, instead, above constructions are corresponding to the gaps between two successive operators in numerical conformal bootstrap. Results obtained from analytical conformal bootstrap approaches suggest that one can obtain sufficient conditions, at least for WF fixed points,  to uniquely determine the specific CFT from single correlator. However, in numerical conformal bootstrap, we do not have a systematical control on the spectra construction. It turns into subtle on how to impose gaps between certain operators. Moreover, the gaps should be guided by the preliminary results from conformal bootstrap.

Interestingly the conserved currents (stress tensor or global symmetry current) and spectra near the unitary bound play critical roles to bootstrap the crossing equation. Applying the constraints that there is a stress tensor $T_{\mu\nu}$ (or conserved global symmetry current $J_\mu$ instead if the theory has global symmetry) and a mild gap for next spin 2 operator $T_{\mu\nu}^\prime$ (or next spin 1 operator $J_\mu^\prime$), we are able to isolate the scaling dimensions of lowest scalar operators in a small closed region. Besides, we will show that if the crossing equation contains extra channels, the spectra near unitary bounds in these channels can also play important roles to bootstrap the crossing equation.

\begin{figure}
\includegraphics[scale=0.63]{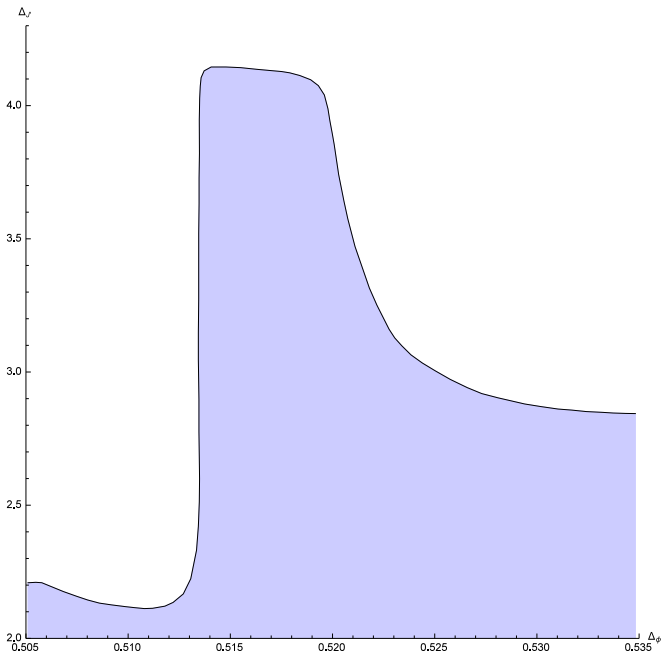}
 \begin{flushright}
\caption{
 Upper bound on the scaling dimension $\Delta_{J^\prime}$ of the second spin 1 operator $J^\prime$ with maximum number of derivative $\Lambda=17$. Here we only require an operator $J^\mu$ saturating the unitary bound while no gap is imposed in the spectra.
 The $(\Delta_{\phi_i}, \Delta_{\phi_i^2})$ is fixed at the bound obtained from single correlator bootstrap with $\Lambda=23$.  } \label{o3pike}
\end{flushright}
\end{figure}

Let us start from the $3D$ $O(3)$ vector model with fundamental representation $\phi_i$.\footnote{This model has been studied in detail in \cite{Kos:2013tga} and  its isolated region has been obtained in \cite{Kos:2015mba}}
For this model, the spectra in spin 1 sector is crucial to generate isolated regions.
The theory admits global $O(3)$ symmetry and correspondingly, there is a spin 1 $O(3)$ conserved current $J_{ij}^\mu(x)$ which appears as the lowest operator ($\Delta_J=2$) in the antisymmetric channel $A^-$ of the $\phi_i\times\phi_j$ OPE:
\be
\phi_i\times\phi_j \sim \sum_{S^+} \delta_{ij} \cO+\sum_{T^+}\cO_{(ij)}+\sum_{A^-}\cO_{[ij]},
\ee
where the $S^+$, $T^+$, $A^-$ denote the $O(N)$ singlet, symmetric and antisymmetric representations, respectively.
As shown in \cite{Kos:2013tga}, from the crossing equation of four point function $\langle\phi_i(x_1)\phi_j(x_2)\phi_k(x_3)\phi_l(x_4) \rangle$ one can get a rigorous upper bound  (Bound \MyRoman{1})  of the dimension of first $O(3)$ singlet scalar $\phi_i^2$ \cite{Kos:2013tga}.
By bootstrapping the same crossing equation with the only requirement that there is a spin 1 operator ($J^\mu$) in $A^-$ channel saturated the unitary bound, we obtain a rigorous upper bound (Figure \ref{o3pike}) on the scaling dimension $\Delta_{J^\prime}$ of next spin 1 operator ${J^\mu_{ij}}^\prime$.
To obtain the bound in Figure \ref{o3pike}, we set $(\Delta_{\phi_i}, \Delta_{\phi_i^2})$ close to the Bound \MyRoman{1}.
The bound on $\Delta_{J^\prime}$ shows a drastic transition near the $O(3)$ fixed point $\Delta_{\phi_i}\simeq0.519$ :
the upper bound has a sharp peak $\Delta_{J^\prime}>4$ in the range $\Delta_{\phi_i}\in(0.513, 0.52)$, while
away from the $O(3)$ fixed point it decreases rapidly.
A similar pike-like upper bound on the second O(N) singlet scalar $\epsilon^\prime$ can also be obtained with constraint $\Delta_{J^\prime}\geqslant3.2$. The upper bound peaks near the $O(3)$ fixed point with $\Delta_{\epsilon^\prime}\sim3.8$.
Similar transition has been observed in \cite{ElShowk:2012ht} for the 3D Ising model in the spin 2 sector.
While for CFTs with global symmetry, it is the global symmetry conserved current
instead of the stress tensor becomes crucial near the fixed point.
The pike-like bound for 3D Ising model has been shown in \cite{Kos:2014bka} for the upper bound on the scaling dimension of next $Z_2$ odd operator $\Delta_{\epsilon^\prime}$ by bootstrapping the mixed correlators, which is the key to obtain isolated region for the scaling dimensions of the lowest $Z_2$ odd and even operators $(\Delta_\sigma,\Delta_\epsilon)$.
Similarly, by assuming there is only one relevant $O(N)$ singlet scalar and a mild gap $\Delta_{J^\prime}>3.5$,\footnote{We can also adopt the gap $\Delta_{J^\prime}>3$, i.e., there is only one relevant spin 1 operator. However, by choosing $\Delta_{J^\prime}>3.5$, which is also expected to be physical from Figure \ref{o3pike},  the bootstrap process gives better estimations on the CFT data.} the scaling dimensions ($\Delta_{\phi_i}, \Delta_{\phi_i^2}$) are limited in a small island.

We have shown that for non-supersymmetric CFTs, constraints from spectra near the unitary bound in certain sectors are strong enough to generate isolated regions of the CFT data.
It is interesting to apply this method to SCFTs which contain more channels in the crossing equation. The spectra near the unitary bounds in these channels also have important effects on solving the crossing equation.
We consider the supersymmetric generalizations of 3D Ising model both with two ($N=1$) and four supercharges ($N=2$). The 3D Ising model, together with its
generalizations with global symmetry and supersymmetry, provide an interesting laboratory for conformal bootstrap. Identification of these CFTs would be a preliminary attempt of the rather ambitious aim of conformal bootstrap on classifying the CFT landscape.

\begin{figure}
\includegraphics[scale=0.7]{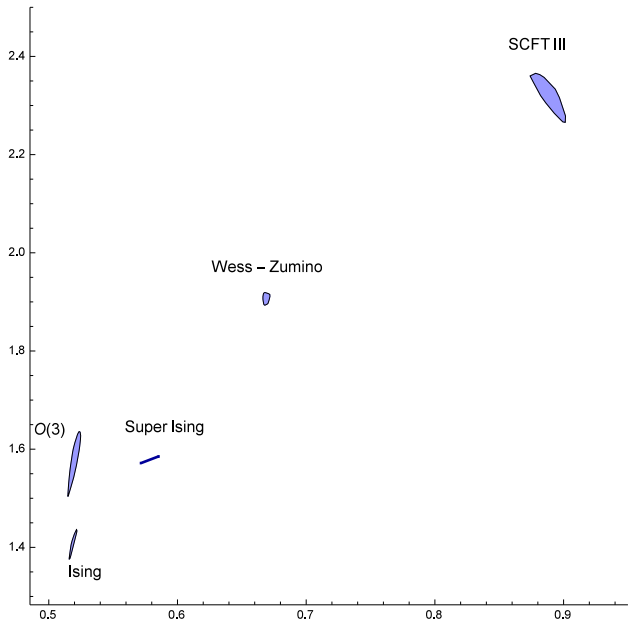}
 \begin{flushright}
\caption{
Archipelago of the 3D CFTs. From the lower left to the upper right, the isolated regions (islands or segment) are corresponding to the 3D Ising model, $O(3)$ vector model, $N=1$ supersymmetric Ising model, Wess-Zumino model and the 3D analogy of $4D$ $N=1$ minimal SCFT. } \label{3dislands}
\end{flushright}
\end{figure}

We show the ``archipelago" of 3D CFTs in Figure \ref{3dislands}. Constraints for these isolated regions are provided in Table 1.
Islands related to the 3D Ising model and $O(3)$ vector model are similar to the seminal results in \cite{Kos:2014bka, Kos:2015mba}, in which the islands are obtained by bootstrapping mixed correlators. In the following parts we will explain the details on the SCFTs presented in Figure \ref{3dislands}.

\vspace{5mm}

{\textbf{3D $N=1$ Supersymmetric Ising Model:}}~~~~~ The 3D $N=1$ supersymmetric Ising model actually shares the same bootstrap equation with the non-supersymmetric case: due to the $Z_2$ symmetry, there is always only one component in any supermultiplet $\Psi=\psi+\chi\theta+\varphi\theta^2$ that may appear in the OPE of two scalars $\phi\times\phi\sim\Psi$  \cite{Bashkirov:2013vya}. For this reason one cannot get specific information on 3D $N=1$ supersymmetric Ising model by bootstrapping the correlators of scalars directly. To resolve this problem, one can bootstrap the correlators with fermions \cite{Iliesiu:2015qra} and the supersymmetry appears at the IR fixed point emergently; or alternatively, one can introduce constraints on the spectra to bootstrap the four point correlator of the lowest $Z_2$ odd scalar $\sigma$:  $\langle\sigma(x_1)\sigma(x_2)\sigma(x_3)\sigma(x_4)\rangle$ \cite{Bashkirov:2013vya}.
Specifically we apply the constraint $\Delta_{\sigma^2}=\Delta_\sigma+1$ obtained from supersymmetry and equation of motion. By introducing extra gaps shown in Table \ref{table} , we obtain a closed interval for $\Delta_\sigma\in(0.571, 0.586)$.

\begin{table}
%Table \MyRoman{1}. Constraints imposed for the 3D CFT archipelago shown in Figure \ref{3dislands}.
\begin{tabular}{l|lllll}
\hline
\hline
                    & ~~Scalar ~~& ~Spin 1~~ & ~Spin 2~~ & ~~non-BPS~  & ~~ $\Lambda$~~   \\
\hline
Ising ~~           &    ~~3          &  ~NA~ & ~S+4.5~ & ~~NA~~& ~~~13   \\
O(3)~~             &    ~~3         &  ~S+3.5~ & ~UB~ & ~~NA~~&  ~~~13 \\
N=1 Ising~~        & ~~3.7           &  ~NA~ & ~S+3.4~ & ~~NA~~&  ~~~30  \\
WZ~~               &        ~~3     &  ~S+3.5~ & ~UB~ & ~~UB~~&  ~~~15  \\
SCFT \MyRoman{3}~~ & ~~4              &  ~S+3.2~ & ~UB~ & ~~S+4~~& ~~~25  \\
\hline
\hline
\end{tabular}\label{table}
\caption{Constraints on the global symmetric invariant scalar, spin 1 and spin 2 sectors imposed in Figure \ref{3dislands}.
Numbers in the ``Scalar" column give the lower bound on the scaling dimension of next scalar in this sector.
``S+$x$" denotes there is an operator $\cO$ saturates the unitary bound and the scaling dimension of next operator $\cO^\prime$ is not smaller than $x$. ``UB" indicates unitary bound and ``NA" means there is no such sector in the crossing equation. The maximum derivatives are given in the column $\Lambda$.}
\end{table}

\vspace{5mm}

{\textbf{3D Wess-Zumino Model:}}~~~~~
The 3D $N=2$ superconformal bootstrap is more illustrative. The $N=2$ extended supersymmetry introduces several channels in the $\Phi\times\Phi$ OPE. The operators  in these channels perform differently near the unitary bounds, which may
relate to different kinks. Based on this property, by applying suitable constraints we are able to isolate multiple islands from the same crossing equation.

The supersymmetric Ising model with four supercharges ($N=2$) is made of a chiral multiplet  $\Phi=\phi+\theta\psi+\cdots$, where the components $\phi$ and $\psi$ are complex scalar and Dirac fermions.
3D $N=2$ SCFTs with a chiral multiplet $\Phi$ have been studied in \cite{Bobev:2015jxa} by bootstrapping the single correlator $\langle\Phi\Phi^\dagger\Phi\Phi^\dagger\rangle$. Details on the superconformal partial wave expansion of this four point correlator and its crossing equation are provided in \cite{Bobev:2015jxa}. The OPE in the $\Phi\Phi$ channel contains non-vanishing $U(1)_R$ charge and is important for our next analysis
\be
\Phi\times\Phi\sim \Phi^2+\sum_{\ell=2,4,\cdots}\bar{Q}\cO_\ell+\bar{Q}^2\cO^*+\sum_{\cO}\bar{Q}^2\cO, \label{PPope}
\ee
where $\Phi^2$ is chiral and $\bar{Q}_{\dot{\alpha}}\cO^{\dot{\alpha}\cdots}_\ell=0$. The anti-chiral term $\cO^*$ appears in the $\Phi\times\Phi$ OPE for theories in $D<4$ while
decouples in 4D $N=1$ SCFTs \cite{Poland:2010wg, Poland:2011ey}.
The last term contains non-protected operators $\cO$ with unitary bound
\be
\Delta_{\cO}\geqslant|2\Delta_{\Phi}-2|+\ell+2.
\ee

\begin{figure}
\includegraphics[scale=0.56]{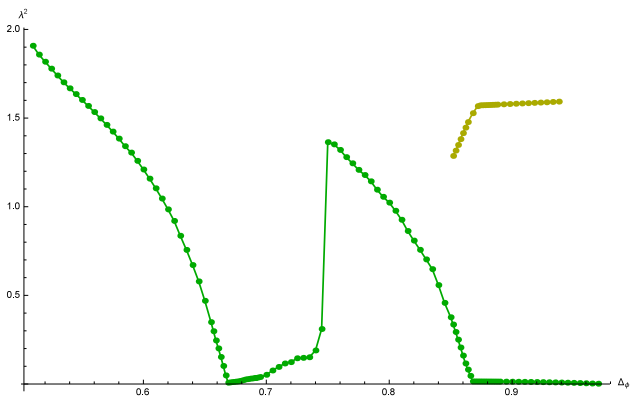}
 \begin{flushright}
\caption{
OPE coefficients ($\lambda^2$) of the operator $\Phi^2$ (green) and the first scalar $\cO_S$ in the non-BPS channel (yellow). $\lambda_{\Phi^2}$ vanishes near the Wess-Zumino point and the SCFT \MyRoman{3}. Besides, there is a jump in $\lambda_{\Phi^2}$ at $\Delta_\Phi=3/4$. $\lambda_{\cO_S}$ shows a clear kink near SCFT \MyRoman{3} and more importantly, it is non-vanishing. The coefficients are estimated with $\Lambda=21$.} \label{opec}
\end{flushright}
\end{figure}

Upper bound
on $\Delta_{\Phi\Phi^\dagger}$ can be obtained through bootstrapping the four-point function $\langle\Phi\Phi^\dagger\Phi\Phi^\dagger\rangle$,
 and it shows three apparent kinks \cite{Bobev:2015jxa}. The first kink is expected to correspond to the Wess-Zumino model and the third kink is the 3D analogy of the 4D minimal SCFT firstly observed in  \cite{Poland:2011ey}, while it is still unclear if the second kink relates to an interacting and unitary SCFT. It is quite interesting to isolate and uniquely determine the putative CFTs with few mild assumptions on the spectrum in certain sectors.

It is shown in Figure \ref{opec} that the OPE coefficient $\lambda_{\Phi^2}$ ($\Phi\times\Phi\sim\lambda_{\Phi^2}\Phi^2$) vanishes near the Wess-Zumino point and SCFT \MyRoman{3}, which indicates the chiral ring condition
$\Phi^2=0$ \cite{Bobev:2015jxa, Poland:2015mta}. Besides, similar to the non-supersymmetric Ising model, we apply the constraint that there is only one relevant $Z_2$ even superconformal primary operator. For the 3D N=2 SCFTs, the stress tensor is not a superconformal primary operator, instead, it is a superdescendent of superconformal multiplet $J_\mu$. The lowest component of $J_\mu$ is the spin 1 $U(1)_R$ symmetry current. In the bootstrap conditions, we require there is such an operator that saturates the spin 1 unitary bound and there is a gap for next spin 1 superconformal primary operator $J^\prime_\mu$: $\Delta_{J^\prime}\geqslant3.5$.
The isolated island in Figure \ref{3dislands} shows the scaling dimensions $(\Delta_\Phi, \Delta_{\Phi\bar{\Phi}})=(0.6678(13), 1.903(10))$ with maximum derivative $\Lambda=15$, which relates to the 3D Wess-Zumino model -- the $N=2$ supersymmetric generalization of the 3D Ising model.

The 3D  Wess-Zumino model contains a single chiral superfield $\Phi=\phi+\theta\psi+\cdots$ with cubic superpotential $W=\lambda \Phi^3$. The Lagrangian in terms of the components is
\bea
\cL_{WZ}=\partial_\mu\bar{\phi}\partial^\mu\phi+i\bar{\psi}\slashed{\partial}\psi &+& \lambda^2(\phi\phi^\dagger)^2+ \nn\\
&&(\lambda\phi\psi_\alpha\epsilon^{\alpha\beta}\psi_\beta+c.c.).
\eea
The cubic superpotential introduces a chiral ring condition $\Phi^2=0$, consistent with the results from superconformal bootstrap.
The chiral superfield $\Phi$ has R-charge $R_\Phi=\frac{2}{3}$, and its scaling dimension $\Delta_\Phi$ is fixed by the $U(1)_R$ charge
\be
\Delta_{\Phi}=|R_\Phi|.
\ee
 The scaling dimension of the operator $\Phi\Phi^\dagger$ has been estimated based on the
$4-\epsilon$ expansion up to the order $O(\epsilon^3)$ \cite{Thomas:2005, Zerf:2016fti, Fei:2016sgs}. Using Pad\'{e} extrapolation it gives $\Delta_{\Phi\Phi^\dagger}\simeq1.909$ \cite{Fei:2016sgs}. The point $(2/3, 1.909)$ just locates in the island ``Wess-Zumino" in Figure \ref{3dislands}. Taking higher order of derivatives $\Lambda=29$,
the island shrinks significantly. The specific slice with $\Delta_\Phi=2/3$ gives a tight range $\Delta_{\Phi\Phi^\dagger}\in (1.9073, 1.9093)$.

\begin{figure}
\includegraphics[scale=0.56]{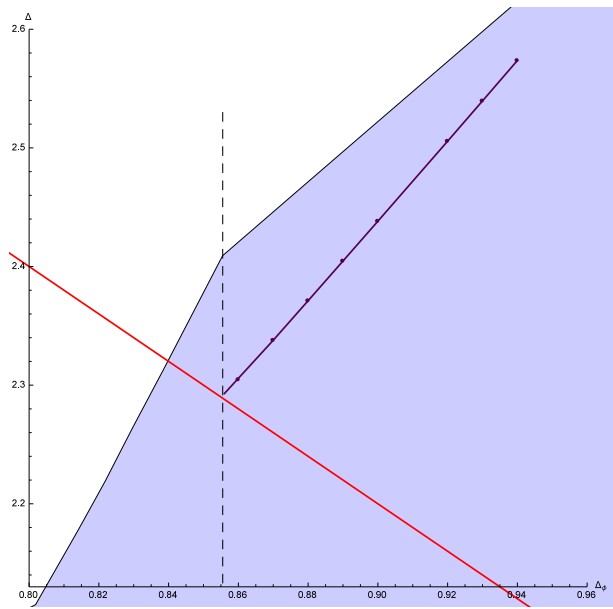}
 \begin{flushright}
\caption{
 Shadowed region gives the upper  bound of scaling dimension $\Delta_{\Phi\Phi^\dagger}$ obtained from single correlator bootstrap ($\Lambda=11$) without assuming any gap in the spectrum. Red line gives the unitary bound of scalars in the non-BPS channel: $\Delta\geqslant|2\Delta_\Phi-2|+2$. The purple line is the scaling dimension $\Delta_{\cO_S}$ obtained from EFM. } \label{bps}
\end{flushright}
\end{figure}

\vspace{5mm}

{\textbf{SCFT \MyRoman{3}:}}~~~~~
The third kink in the upper bound on $\Delta_{\Phi\Phi^\dagger}$ is more subtle. Apparently only the constraints used for the first kink is not sufficient and we need extra condition in the bootstrap setup. Actually it corresponds to the spectra in the non-BPS channel in the $\Phi\times\Phi$ OPE.
In Figure \ref{bps} we show the evolution of the scaling dimension $\Delta_\cO$ of operator $\cO$ (purple line), which is the first scalar in the non-BPS channel in (\ref{PPope}). The scaling dimension $\Delta_\cO$ is estimated using extremal functional method (EFM) \cite{ElShowk:2012hu}, and it hits the unitary bound on non-BPS channel at the point related to SCFT \MyRoman{3}. Its OPE coefficient $\lambda_\cO$ ($\Phi\times\Phi\sim\lambda_\cO\cO$) is still non-vanishing as shown in the Figure \ref{opec}. Therefore this operator does not decouple from the theory. The island suggests a new SCFT with a short multiplet $\cO$ saturating the unitary bound on non-BPS channel. Similar property also holds for the $4D$ minimal SCFT \cite{inpre}. Likewise, to obtain better estimation on the CFT data of SCFT \MyRoman{3}, we introduce the constraints in the non-BPS sector of $\Phi\times\Phi$ channel: i, there is an operator $\cO_S$ saturates the unitary bound, i.e., $\Delta_{\cO_S}=|2\Delta_\Phi-2|+2$; ii, there is a gap between $\Delta_{\cO_S}$ and the next scalar in this sector $\cO^\prime$. Interestingly, with these extra constraints the third kink can be further limited to a small island, as shown in Figure \ref{3dislands}. Without the saturation condition, the isolated region disappears even by assuming that the scaling dimension of the first scalar in the non-BPS channel is slightly above unitary bound.

\vspace{5mm}

The CFTs we isolated above are already shown as kinks in the  bounds of scaling dimensions of certain operators.\footnote{The 3D $N=1$ supersymmetric Ising model is slightly different. It does not relate to any kink in the bound from single scalar correlator bootstrap. In this case an extra constraint $\Delta_{\phi^2}=\Delta_\phi+1$ is needed to obtain a closed range of $\Delta_\phi$.} In general one would expect the appearance of kink reflects certain irregular behaviors of the spectra, like certain lower spectrum becomes null or hits the unitary bound, and decouple from the theory on one side of the kink, which makes the spectra evolving discontinuously across the kink. By introducing saturation condition and gaps in the spectra, we are able to capture the traits of the kinks and provide stronger constraints on the CFT data.

In principal one expects to obtain better estimation on the CFT data with more constraints. However, it is quite surprising that these few mild assumptions on the spectra, especially the saturation condition from stress tensor or conserved global symmetry current, can lead to a small isolated set of solutions of the crossing equation.
Optimistically the set of isolated solutions is expected to converge to a point which gives the exact CFT data. In this sense, our results suggest such kind of CFTs are actually uniquely determined by its few lowest operators. Moreover, as shown in the examples on 3D $N=2$ SCFTs, with $N=2$ supersymmetry we have both BPS and non-BPS channels in the crossing equation. Besides the well-known Wess-Zumino model, a putative SCFT appears when the unitary bound in non-BPS channel is saturated by an operator.

Previously the 3D Ising model (or 3D/5D $O(N)$ vector models) is isolated by imposing gaps in both $Z_2$ even and odd sectors ($O(N)$ singlet and vector representations). To get access to the $Z_2$ odd ($O(N)$ vector) sector, one has to work with mixed correlators.
Here we show that the conditions for isolated solutions can be realized in single correlator, which is simpler and also quite general. Since the calculation load for single correlator bootstrap is much less than that with mixed correlators, we expect our method can provide precise estimation on the CFT data, especially for the 3D $O(N)$ vector models \cite{Kos:2015mba, Kos:2016ysd}. The gaps on scaling dimensions introduced in the bootstrap conditions are important for the efficiency in estimating CFT data. It would be helpful if we can find a way to optimize the gaps.

The examples on 3D $N=2$ SCFTs suggest a likely connection between the ``kink" CFTs and the spectra near the unitary bound in certain channels. This putative connection needs to be tested by the CFTs with more channels in the crossing equation. In particular, this scenario could be quite interesting for SCFTs with extended supersymmetry, which contain various kinds of BPS sectors as well independent structures in the crossing equations. They have been studied using conformal bootstrap, such as the 3D $N=8$ SCFTs \cite{Chester:2014fya}, 4D $N=2$ and $N=4$ SCFTs \cite{Beem:2013qxa, Beem:2014zpa} as well as the 6D $(2, 0)$ SCFTs \cite{Beem:2015aoa} which have no classical Lagrangian description. Hopefully we can carve out more constrained space on these SCFTs, which may be classified based on their spectra near certain channels in the crossing equation.

We can also test this possible connection by bootstrapping the single correlator of spinning operators. Spinning operator bootstrap has been done for 3D fermions \cite{Iliesiu:2015qra, Iliesiu:2017nrv} and 3D conserved global symmetry currents \cite{Dymarsky:2017xzb}. Four point correlators with spinning operators include many independent tensor structures. One may expect the spectra near the unitary bounds in different channels relate to different CFTs, and by imposing assumptions on the spectra near the unitary bound in certain channels, we may get access to the relevant CFTs.

We have shown that the lowest spectra in the stress tensor or conserved global symmetry current sector are powerful in carving out the CFT space, and it is expected to be helpful to study/classify more complex and abundant CFTs. However, it is quite puzzle why these spectra are so useful in conformal bootstrap when combined with crossing symmetry and unitarity.
It would be very important to understand the analytical reasons behind these numerical results. A promising attempt on analytical approach of the conformal bootstrap is initiated in \cite{Mazac:2016qev}.
We expect this question can be partially clarified by analyzing the special roles of these lowest spectra and the gaps in relevant sectors in constructing the analytical extremal function of the crossing equation.

\noindent {\bf Acknowledgements. }
We are grateful to the valuable discussions with Miguel Costa, Rajesh Gopakumar, Song He,   Dalimil Mazac, David Meltzer, Miguel Paulos, Joao Penedones, David Polland, Christopher Pope, Daniel Robbins, Junchen Rong,  and David Simmons-Duffin. ZL is grateful to Christopher Pope for his encourage and support.
ZL would also like to thank ICTP-SAIFR S\~{a}o Paulo and TASI 2017 for the hospitality during the completion of this work. The work of ZL is supported by the DOE grant DE-FG02-13ER42020.
The work of NS is supported by ITP-CAS. 
The computations in this paper were run on HPC Cluster of SKLTP/ITP-CAS  and on the Mac Lab cluster supported by the Department of Physics and Astronomy,  Texas A\&M University.

\end{document}